\documentclass[12pt]{article}

\usepackage{enumitem}
\usepackage{float}
\usepackage{cite}
\usepackage{amsfonts}
\usepackage{amssymb}
\usepackage{amsmath}
\usepackage{xcolor}
\usepackage{mathtools}

\usepackage{graphicx}

\def\gtwid{\mathrel{\raise.3ex\hbox{$>$\kern-.75em\lower1ex\hbox{$\sim$}}}}
\def\ltwid{\mathrel{\raise.3ex\hbox{$<$\kern-.75em\lower1ex\hbox{$\sim$}}}}
\def\square{\kern1pt\vbox{\hrule height 1.2pt\hbox{\vrule width 1.2pt\hskip 3pt
   \vbox{\vskip 6pt}\hskip 3pt\vrule width 0.6pt}\hrule height 0.6pt}\kern1pt}

\begin{document}

\begin{titlepage}

\begin{flushright}
UFIFT-QG-24-09
\end{flushright}

\vskip 2cm

\begin{center}
{\bf Resumming Fermion Loops for Inflationary Gravity}
\end{center}

\vskip 0.5cm

\begin{center}
A. J. Foraci$^{*}$ and R. P. Woodard$^{\dagger}$
\end{center}

\vskip 0.5cm

\begin{center}
\it{Department of Physics, University of Florida,\\
Gainesville, FL 32611, UNITED STATES}
\end{center}

\vspace{0.5cm}

\begin{center}
ABSTRACT
\end{center}
We compute the 1-loop contribution to the graviton self-energy from a 
loop of massless fermions on a general cosmological background. The
result is used to quantum-correct the linearized Einstein equation on
de Sitter background and work out 1-loop corrections to gravitational
radiation and to the response to a point mass. The renormalization
group is employed to sum these to all orders for as long as the de
Sitter phase persists.

\begin{flushleft}
\end{flushleft}

\vskip 2cm

\begin{flushleft}
$^{*}$ e-mail: aforaci@ufl.edu \\
$^{\dagger}$ e-mail: woodard@phys.ufl.edu
\end{flushleft}

\end{titlepage}

\section{Introduction}

The inflationary production of long wavelength gravitons 
\cite{Starobinsky:1979ty} induces large logarithmic corrections 
to the kinematics and long range forces carried by other fields
\cite{Miao:2006gj,Glavan:2013jca,Wang:2014tza,Glavan:2021adm}. Even 
if the particles associated with these other particles are not 
produced during inflation, the redshift of their vacuum energies can 
still cause secular changes in gravitational radiation and in the 
force of gravity \cite{Wang:2015eaa,Miao:2024atw,Foraci:2024vng}. 
This paper is devoted to working out the changes induced by a loop 
of massless fermions.

Section 2 computes the 1-loop contribution to the graviton self-energy
from a loop of massless fermions on any cosmological background, 
slightly extending an old calculation on flat space background 
\cite{Capper:1973mv}. In section 3 we use this result to 
quantum-correct the linearized Einstein equation on de Sitter 
background. Solving this equation gives us 1-loop corrections to 
gravitational radiation and to the response to a static point mass. We
also show how a variant of the renormalization group can be used to 
resum these results. Our conclusions comprise section 4.

\section{The Graviton Self-Energy from Fermions}

\begin{figure}[H]
\centering
\vskip 1cm
\includegraphics[width=8cm]{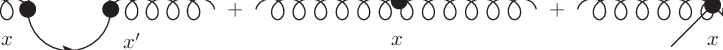}
\caption{\footnotesize Fermionic contributions to the 1-loop graviton 
self-energy. Solid lines stand for fermions and curly lines for gravitons.}
\label{diagrams}
\end{figure}

The 1PI (one-particle-irreducible) 2-point function for the graviton is 
known as the graviton self-energy $-i [\mbox{}^{\mu\nu} 
\Sigma^{\rho\sigma}](x;x')$. The diagrams representing the 1-loop fermion 
correction to it are shown in Figure~\ref{diagrams}. 

The background geometry of ($D$-dimensional) cosmology is,
\begin{equation}
ds^2 = -dt^2 + a^2(t) d\vec{x} \!\cdot\! d\vec{x} = a^2 [-d\eta^2 + d\vec{x}
\!\cdot\! d\vec{x} ] \; . \label{background}
\end{equation}
Here $t$ is the co-moving time and $\eta$ is the conformal time. We define 
the graviton field $h_{\mu\nu}(x)$ by conformally transforming the full
metric,
\begin{equation}
g_{\mu\nu}(x) \equiv a^2 \widetilde{g}_{\mu\nu} \equiv a^2 [\eta_{\mu\nu} +
\kappa h_{\mu\nu}(x)] \qquad , \qquad \kappa^2 \equiv 16 \pi G \; .
\label{graviton}
\end{equation}
Its indices are raised and lowered using the Minkowski metric $h^{\mu}_{~\nu}
\equiv \eta^{\mu\rho} h_{\nu\rho}$. 

The massless Dirac Lagrangian for a general metric is,
\begin{equation}
\mathcal{L}_{\rm Dirac} = \overline{\psi} e^{\mu}_{~a} \gamma^a \Bigl(i 
\partial_{\mu} \!-\! \tfrac12 A_{\mu bc} J^{bc} \Bigr) \psi \sqrt{-g} \; . 
\label{Dirac}
\end{equation}
Here $\gamma^a$ are the gamma matrices, $e^{\mu}_{~a}(x)$ is the vierbein 
field, $A_{\mu bc}(x)$ is the spin connection and the $J^{bc}$ are spin
generators,
\begin{equation}
g^{\mu\nu} = e^{\mu}_{~a} e^{\nu}_{~b} \eta^{ab} \quad , \quad 
A_{\mu bc} \equiv e^{\nu}_{~b} (e_{\nu c , \mu} - \Gamma^{\rho}_{~\mu\nu}
e_{\rho c}) \quad , \quad J^{bc} \equiv \tfrac{i}{4} [\gamma^{b}, 
\gamma^{c}] \; . \label{spinstuff}
\end{equation}
If the local Lorentz gauge is fixed by requiring the vierbein to be 
symmetric $e_{\mu a} = e_{a \mu}$ there are no local Lorentz ghosts and 
one can regard the vierbein as a function of the graviton field 
\cite{Woodard:1984sj},
\begin{equation}
e_{\mu b} = a \Bigl[\sqrt{\widetilde{g} \eta^{-1}} \,\Bigr]_{\mu}^{~c}
\times \eta_{c b} = a \Bigl[\eta_{\mu b} + \tfrac{\kappa}{2} h_{\mu b} - 
\tfrac{\kappa^2}{8} h_{\mu}^{~ c} h_{c b} + \dots\Bigr] . \label{vierbein}
\end{equation}

Because $\mathcal{L}_{\rm Dirac}$ is conformally invariant for any
dimension $D$, it has no dependence on the scale factor when both the 
fermion and the metric are conformally rescaled,
\begin{equation}
\psi \equiv a^{\frac{D-1}{2}} \widetilde{\psi} \qquad , \qquad 
\mathcal{L}_{\rm Dirac} = \overline{\widetilde{\psi}} \widetilde{e}^{\mu}_{~a} 
\gamma^a \Bigl(i \partial_{\mu} \!-\! \tfrac12 \widetilde{A}_{\mu bc} J^{bc} 
\Bigr) \widetilde{\psi} \sqrt{-\widetilde{g}} \; . \label{newDirac}
\end{equation}
This means that dependence on the scale factor can only enter the 
graviton self-energy through counterterms. Neither Einstein-Maxwell 
\cite{Deser:1974cz,Deser:1974zzd} nor Einstein-Dirac \cite{Deser:1974cy} 
is perturbatively renormalizable, even at 1-loop order, but the 
divergences of any theory can be subtracted using BPHZ counterterms 
(Bogoliubov, Parasiuk \cite{Bogoliubov:1957gp}, Hepp \cite{Hepp:1966eg} 
and Zimmermann \cite{Zimmermann:1968mu,Zimmermann:1969jj}). The ones 
needed to renormalize any matter loop contribution to the graviton 
self-energy take the form \cite{tHooft:1974toh},
\begin{equation}
\Delta \mathcal{L}_{\rm Einstein} = c_1 R^2 \sqrt{-g} + c_2 
C^{\alpha\beta\gamma\delta} C_{\alpha\beta\gamma\delta} \sqrt{-g} \; .
\label{DeltaEinstein}
\end{equation}

The fermion propagator from (\ref{newDirac}) can be expressed in terms of
the massless scalar propagator on flat space $i\Delta(x;x')$,
\begin{equation}
\Bigl\langle \Omega_0 \Bigl\vert T[\widetilde{\psi}_i(x) 
\overline{\widetilde{\psi}}_j(x') ] \Bigr\vert \Omega_0 \Bigr\rangle = i
\gamma^{\mu}_{ij} \partial_{\mu} i\Delta(x;x') = -\tfrac{\Gamma(\frac{D}2)}{
2 \pi^{\frac{D}2}} \tfrac{ \gamma^{\mu}_{ij} \Delta x_{\mu}}{\Delta x^D} \; .
\label{propagator} 
\end{equation}
Here $\Delta x^2 \equiv \Vert \vec{x} - \vec{x}' \Vert^2 - (\vert\eta - 
\eta'\vert - i \epsilon)^2$ is the conformal coordinate interval. The 
3-point vertex is,
\begin{equation}
\tfrac{i \kappa}{2} h_{\mu\nu} \Bigl[\eta^{\mu\nu} \overline{\widetilde{\psi}}
i \gamma^{\alpha} \partial_{\alpha} \widetilde{\psi} - \overline{\widetilde{\psi}}
i \gamma^{\mu} \partial^{\nu} \widetilde{\psi} + \partial_{\alpha} \Bigl(
\overline{\widetilde{\psi}} \gamma^{\mu} J^{\nu \alpha} \widetilde{\psi} \Bigr)
\Bigr] \; . \label{3point}
\end{equation} 
Note that the final term involving the spin generator can be written as,
\begin{equation}
\gamma^{(\mu} J^{\nu) \alpha} = \tfrac{i}{2} \eta^{\alpha (\mu} \gamma^{\nu)}
- \tfrac{i}{2} \eta^{\mu\nu} \gamma^{\alpha} \; , \label{simplification}
\end{equation}
where parenthesized indices are symmetrized. 

Because the coincidence limit of the propagator (\ref{propagator}) vanishes 
in dimensional regularization the middle diagram of Figure~\ref{diagrams} 
vanishes and we do not need the 4-point interaction. Because acting $i 
\gamma^{\mu} \partial_{\mu}$ on the fermion propagator gives a delta function 
we require only two of the 3-point vertices,
\begin{equation}
V_1 \equiv h_{\mu\nu} \times \tfrac{\kappa}{2} \overline{\widetilde{\psi}}
\gamma^{\mu} \partial^{\nu} \widetilde{\psi} \qquad , \qquad V_2 \equiv 
h_{\mu\nu} \times -\tfrac{\kappa}{4} \partial^{\mu} [\overline{\widetilde{\psi}} 
\gamma^{\nu} \widetilde{\psi}] \; . \label{V1V2}
\end{equation}
The primitive contribution to the graviton self-energy is the sum of products 
of one of these vertices at $x^{\mu}$ and another at ${x'}^{\mu}$. The first 
such product is,
\begin{eqnarray}
\lefteqn{ -i [\mbox{}^{\mu\nu} \Sigma^{\rho\sigma}_{11}](x;x') =
\tfrac{\kappa^2}{4} \Bigl\langle \Omega_0 \Bigl\vert T\Bigl[ 
\overline{\widetilde{\psi}}(x) \gamma^{(\mu} \partial^{\nu)} 
\widetilde{\psi}(x) \times \overline{\widetilde{\psi}}(x') \gamma^{(\rho} 
{\partial'}^{\sigma)} \widetilde{\psi}(x') \Bigr] \Bigr\vert \Omega_0 
\Bigr\rangle } \nonumber \\
& & \hspace{0cm} =-\tfrac{\kappa^2 \Gamma^2(\frac{D}2)}{16 \pi^D} {\rm Tr}\Bigl[
\gamma^{(\rho} {\partial'}^{\sigma)} (\tfrac{\gamma^{\beta} \Delta x_{\beta}}{
\Delta x^D}) \gamma^{(\mu} \partial^{\nu)} (\tfrac{\gamma^{\alpha} \Delta 
x_{\alpha}}{\Delta^D}) \Bigr] \; , \qquad \\
& & \hspace{0cm} = -\tfrac{\kappa^2 \Gamma^2(\frac{D}2)}{4 \pi^D} \Bigl\{
-\tfrac{\eta^{\mu\nu} \eta^{\rho\sigma}}{\Delta x^{2D}} + \tfrac{D [
\eta^{\mu\nu} \Delta x^{\rho} \Delta x^{\sigma} + \Delta x^{\mu} \Delta x^{\nu}
\eta^{\rho\sigma}]}{\Delta x^{2D+2}} \nonumber \\
& & \hspace{5.3cm} + \tfrac{D^2 \Delta x^{(\mu} \eta^{\nu) (\rho} \Delta 
x^{\sigma)}}{\Delta x^{2D + 2}} - \tfrac{2 D^2 \Delta x^{\mu} \Delta x^{\nu} 
\Delta x^{\rho} \Delta x^{\sigma}}{\Delta x^{2D+4}} \Bigr\} \; . \qquad 
\label{11term}
\end{eqnarray}
The other three contributions are,
\begin{eqnarray}
-i [\mbox{}^{\mu\nu} \Sigma^{\rho\sigma}_{12}](x;x') &\!\!\! = \!\!\!&
-\tfrac{\kappa^2 \Gamma^2(\frac{D}2)}{8 \pi^D} \partial^{(\rho} \Bigl\{ 
\tfrac{\Delta x^{\sigma)} \eta^{\mu\nu} + D \eta^{\sigma) (\mu} 
\Delta x^{\nu)}}{\Delta x^{2D}} - \tfrac{2 D \Delta x^{\sigma)} \Delta x^{\mu} 
\Delta x^{\nu}}{\Delta x^{2D+2}} \Bigr\} \; , \qquad \label{12term} \\
-i [\mbox{}^{\mu\nu} \Sigma^{\rho\sigma}_{21}](x;x') &\!\!\! = \!\!\!&
-\tfrac{\kappa^2 \Gamma^2(\frac{D}2)}{8 \pi^D} \partial^{(\mu} \Bigl\{ 
\tfrac{\Delta x^{\nu)} \eta^{\rho\sigma} + D \eta^{\nu) (\rho} 
\Delta x^{\sigma)}}{\Delta x^{2D}} - \tfrac{2 D \Delta x^{\nu)} 
\Delta x^{\rho} \Delta x^{\sigma}}{\Delta x^{2D+2}} \Bigr\} \; , \qquad 
\label{21term} \\
-i [\mbox{}^{\mu\nu} \Sigma^{\rho\sigma}_{22}](x;x') &\!\!\! = \!\!\!&
-\tfrac{\kappa^2 \Gamma^2(\frac{D}2)}{16 \pi^D} \partial^{(\mu} 
\partial^{((\rho} \Bigl\{ \tfrac{\eta^{\nu) \sigma))}}{\Delta x^{2D -2}}
- \tfrac{2 \Delta x^{\nu)} \Delta x^{\sigma))}}{\Delta x^{2D}} \Bigr\} \; .
\qquad \label{22term} 
\end{eqnarray}

The next step is to extract a total of six derivatives from each term using 
identities of the form,
\begin{eqnarray}
\tfrac{1}{\Delta x^{2D}} &\!\!\! = \!\!\!& \tfrac{\partial^6}{8 D (D-1) 
(D-2)^2 (D-3) (D-4)} (\tfrac{1}{\Delta x^{2D - 6}}) \; , \qquad 
\label{ID1} \\
\tfrac{\Delta x^{\mu} \Delta x^{\nu}}{\Delta x^{2D+2}} &\!\!\! = \!\!\!& 
\Bigl[ \tfrac{\eta^{\mu\nu} \partial^6}{16 D^2 (D-1) (D-2)^2 (D-3) (D-4)}
+ \tfrac{\partial^{\mu} \partial^{\nu} \partial^4}{16 D (D-1) (D-2)^2 (D-3)
(D-4)} \Bigr] \tfrac1{\Delta x^{2D-6}} \; . \qquad \label{ID2} 
\end{eqnarray}
When this is done and all four terms summed, the result can be expressed
using the transverse projection operator,
\begin{equation}
\Pi^{\mu\nu} \equiv \partial^{\mu} \partial^{\nu} - \eta^{\mu\nu} 
\partial^2 \; . \label{transverse}
\end{equation}
In these terms the primitive contribution is,
\begin{equation}
-i [\mbox{}^{\mu\nu} \Sigma^{\rho\sigma}_{\rm prim}] = 
-\tfrac{\kappa^2 \Gamma^2(\frac{D}2)}{64 \pi^D} \Bigl[ \Pi^{\mu (\rho}
\Pi^{\sigma) \nu} - \tfrac{\Pi^{\mu\nu} \Pi^{\rho\sigma}}{D-1} \Bigr]
\tfrac{\partial^2}{(D+1) (D-2)^2 (D-3) (D-4)} (\tfrac1{\Delta x^{2D-6}})
\; . \label{primitive}
\end{equation}
Note that expression (\ref{primitive}) is transverse and traceless for
all $D$.

We must now localize the divergences and subtract them using the 
counterterms (\ref{DeltaEinstein}). Localization is accomplished by
adding zero in the form of the massless scalar propagator equation
\cite{Onemli:2002hr},
\begin{eqnarray}
\tfrac{\partial^2}{D-4} \Bigl[ \tfrac{1}{\Delta x^{2D-6}} \Bigr] &\!\!\! 
= \!\!\!& \tfrac{\partial^2}{D-4} \Bigl[ \tfrac{1}{\Delta x^{2D-6}} -
\tfrac{\mu^{D-4}}{\Delta x^{D-2}} \Bigr] + \tfrac{\mu^{D-4}}{D-4}
\tfrac{4 \pi^{\frac{D}2} i \delta^D(x - x')}{\Gamma(\frac{D}2 - 1)} \; ,
\qquad \\
&\!\!\! = \!\!\!& \tfrac{\mu^{D-4}}{D-4} \tfrac{4 \pi^{\frac{D}2} i 
\delta^D(x - x')}{\Gamma(\frac{D}2 - 1)} - \tfrac{\partial^2}{2} \Bigl[
\tfrac{\ln(\mu^2 \Delta x^2)}{\Delta x^2} \Bigr] + O(D \!-\! 4) \; .
\qquad \label{localdiv}
\end{eqnarray}
Because the primitive result (\ref{primitive}) is transverse and traceless,
the Eddington ($R^2$) counterterm has coefficient zero, $c_1 = 0$. The 
variation of the Weyl ($C^{\alpha\beta\gamma\delta} C_{\alpha\beta\gamma
\delta}$) counterterm is,
\begin{equation}
\tfrac{\delta^2 i \Delta S}{\delta h_{\mu\nu}(x) \delta h_{\rho\sigma}(x')}
\Bigl\vert_{h_{\alpha\beta} = 0} = 2 c_2 \kappa^2 \mathcal{C}^{\alpha\beta
\gamma\delta\mu\nu} \Bigl[ a^{D-4} \mathcal{C}_{\alpha\beta\gamma\delta}^{
~~~~~ \rho\sigma} i \delta^D(x \!-\! x')\Bigr] \; . \label{Weylcterm}
\end{equation}
Here the 2nd order tensor differential operator $\mathcal{C}_{\alpha\beta
\gamma\delta}^{~~~~~ \mu\nu}$ is obtained by expanding the Weyl tensor of 
the conformally transformed metric $\widetilde{g}_{\mu\nu} \equiv 
\eta_{\mu\nu} + \kappa h_{\mu\nu}$,
\begin{equation}
\widetilde{C}_{\alpha\beta\gamma\delta} \equiv \mathcal{C}_{\alpha\beta
\gamma\delta}^{~~~~~\mu\nu} \!\times\! \kappa h_{\mu\nu} + O(\kappa^2 h^2)
\; . \label{Cdef}
\end{equation}
It is manifestly traceless and transverse, and its explicit form can be 
found in \cite{Park:2011ww,Leonard:2014zua}. Of great significance is the
simplification which occurs when acted on a delta function,
\begin{equation}
\mathcal{C}^{\alpha\beta\gamma\delta\mu\nu} \mathcal{C}_{\alpha\beta
\gamma\delta}^{~~~~~ \rho\sigma} \delta^D(x \!-\! x') = (\tfrac{D-3}{D-2})
\Bigl[ \Pi^{\mu (\rho} \Pi^{\sigma) \nu} - \tfrac{\Pi^{\mu\nu} 
\Pi^{\rho\sigma}}{D-1} \Bigr] \delta^D(x \!-\! x') \; . \label{Csimp}
\end{equation}
It follows that the divergences will cancel if we choose,
\begin{equation}
c_2 = \tfrac{\mu^{D-4}}{64 \pi^{\frac{D}2}} \tfrac{\Gamma(\frac{D}2)}{
(D+1) (D-3)^2 (D-4)} \; . \label{c2def}
\end{equation}
The final, renormalized and unregulated result is,
\begin{equation}
-i [\mbox{}^{\mu\nu} \Sigma^{\rho\sigma}_{\rm ren}] = \tfrac{\kappa^2}{
2^8 \cdot 5 \cdot \pi^4} \mathcal{C}^{\alpha\beta\gamma\delta\mu\nu} 
\!\times\! {\mathcal{C}'}_{\alpha\beta\gamma\delta}^{~~~~~ \rho\sigma} 
\Bigl\{4 \pi^2 \ln(a a') i \delta^4(x \!-\! x') + \partial^2 [ 
\tfrac{\ln(\mu^2 \Delta x^2)}{\Delta x^2} ] \Bigr\} . 
\label{renormalized}
\end{equation}

\section{The Gravitational Response to Fermions}

The graviton self-energy can be used to quantum-correct the linearized
Einstein equation,
\begin{equation}
\mathcal{L}^{\mu\nu\rho\sigma} \kappa h_{\rho\sigma}(x) - \int \!\! d^4x'
\, [\mbox{}^{\mu\nu} \Sigma^{\rho\sigma}](x;x') \kappa h_{\rho\sigma}(x')
= \tfrac{\kappa^2}{2} T^{\mu\nu}(x) \; . \label{EinsteinEQN}
\end{equation}
With the stress tensor $T^{\mu\nu}$ set to zero one can study plane wave
gravitons,
\begin{equation}
\kappa h_{\mu\nu} = \epsilon_{\mu\nu} e^{i \vec{k} \cdot \vec{x}} u(t,k) \; ,
\label{gravrad}
\end{equation}
where the transverse-traceless and purely spatial polarization tensor 
$\epsilon_{\mu\nu}$ is the same as on flat space background and the tree 
order mode function is, 
\begin{equation}
u^{\rm tree} = \tfrac{H}{\sqrt{2 k^3}} [1 - \tfrac{i k}{a H} ] 
\exp[\tfrac{i k}{H a}] \label{treemode}
\end{equation}
The gravitational response to a point mass requires two potentials,
\begin{equation}
T^{\mu\nu}(x) = -\delta^{\mu}_{~0} \delta^{\nu}_{~0} M a \delta^3(\vec{x}) 
\qquad , \qquad 
ds^2 = -[1 \!-\! 2 \Psi] dt^2 + a^2 [1 \!-\! 2 \Phi] d\vec{x} \!\cdot\!
d\vec{x} \; . \label{PointMass}
\end{equation}

Employing the in-out self-energy (\ref{renormalized}) in this equation 
would be inappropriate because it is neither real nor causal. Both of
these difficulties can be avoided by employing the in-in, or 
Schwinger-Keldysh, formalism \cite{Chou:1984es,Jordan:1986ug,
Calzetta:1986ey}. The rules for converting an in-out self-energy of the 
form (\ref{primitive}) to in-in form are straightforward 
\cite{Ford:2004wc} and the result is,
\begin{equation}
[\mbox{}^{\mu\nu} \Sigma_{\rm SK}^{\rho\sigma}](x;x') = -
\tfrac{\kappa^2}{2^9 \cdot 5 \cdot \pi^3} \, 
\mathcal{C}^{\alpha\beta\gamma\delta\mu\nu} \!\times\!
{\mathcal{C}'}_{\alpha\beta\gamma\delta}^{~~~~~\rho\sigma} \Bigl[ 8 \pi 
\ln(a a') \delta^4(x \!-\! x') + f_B(x \!-\! x')\Bigr] . \label{SKSE}
\end{equation}
The function $f_B(x - x')$ is \cite{Miao:2024pwd},
\begin{equation}
f_B(x \!-\! x') \equiv \partial^4 \Bigl\{ \theta(\Delta \eta \!-\! \Delta r)
\Bigl(\ln[\mu^2 (\Delta \eta^2 \!-\! \Delta r^2)] \!-\! 1 \Bigr) \Bigr\} 
\; , \label{fB}
\end{equation}
where $\Delta \eta \equiv \eta - \eta'$ and $\Delta r \equiv \Vert \vec{x} -
\vec{x}'\Vert$.

Our result (\ref{SKSE}) is valid for an arbitrary cosmological background
(\ref{background}). To facilitate actually solving equation (\ref{EinsteinEQN})
we will specialize to de Sitter, whose scale factor is $a = e^{H t} = 
-(H\eta)^{-1}$, with Hubble constant $H$. On this background the Lichnerowicz 
operator is,
\begin{eqnarray}
\lefteqn{\mathcal{L}^{\mu\nu\rho\sigma} h_{\rho\sigma} = \tfrac12 a^2 
\Bigl[\partial^2 h^{\mu\nu} \!-\! \eta^{\mu\nu} \partial^2 h \!+\! 
\eta^{\mu\nu} \partial^{\rho} \partial^{\sigma} h_{\rho\sigma} \!+\! 
\partial^{\mu} \partial^{\nu} h \!-\! 2 \partial^{\rho} 
\partial^{(\mu} h^{\nu)}_{~~\rho} \Bigr] } \nonumber \\
& & \hspace{1cm} + a^3 H \Bigl[ \eta^{\mu\nu} \partial_0 h \!-\! \partial_0
h^{\mu\nu} \!-\! 2 \eta^{\mu\nu} \partial^{\rho} h_{\rho 0} \!+\! 2 
\partial^{(\mu} h^{\nu)}_{~~0} \Bigr] \!+\! 3 a^4 H^2 \eta^{\mu\nu} h_{00} 
\; . \qquad \label{Lichnerowicz}
\end{eqnarray}
Because the massless Dirac contribution (\ref{SKSE}) is $\tfrac12$ times the
contribution from electromagnetism \cite{Wang:2015eaa}, we can read off the 
1-loop solutions for Dirac from those for Maxwell. Hence the 1-loop 
correction to the electric components of the Weyl tensor for plane wave 
gravitational radiation is \cite{Foraci:2024vng},
\begin{equation}
C_{0i0j}(x) \longrightarrow C_{0i0j}^{\rm tree}(x) \Bigl\{1 \!+\!
\tfrac{\kappa^2 H^2}{80 \pi^2} \, \ln{(a)} + \dots \Bigr\} \; . 
\label{DiracWeyl}
\end{equation}
The 1-loop corrections to the Newtonian potential and the gravitational
slip are \cite{Wang:2015eaa},
\begin{eqnarray}
\Psi(t,r) &\!\!\! = \!\!\!& \tfrac{G M}{a r} \Bigl\{1 + \tfrac{\kappa^2}{
120 \pi^2 a^2 r^2} + \tfrac{\kappa^2 H^2}{80 \pi^2} \ln(a H r) + \dots 
\Bigr\} \; , \qquad \label{DiracNewton} \\
\Psi(t,r) + \Phi(t,r) &\!\!\! = \!\!\!& \tfrac{G M}{a r} \Bigl\{ 0 +
\tfrac{\kappa^2}{240 \pi^2 a^2 r^2} -\tfrac{\kappa^2 H^2}{80 \pi^2} +
\dots \Bigr\} \; . \qquad \label{DiracSlip}
\end{eqnarray}
The fractional $\kappa^2/a^2 r^2$ corrections are de Sitter descendants
of old effects on flat space background \cite{Capper:1973mv}; the terms
proportional to $\kappa^2 H^2$ are new. 

During a prolonged period of de Sitter inflation, the secular 1-loop
corrections in (\ref{DiracWeyl}) and (\ref{DiracNewton}) must eventually 
overwhelm the tree order result. At this point perturbation theory breaks
down and one must invoke a nonperturbative resummation scheme to work out
what happens. The renormalized self-energy (\ref{renormalized}) shows the
close connection between the renormalization scale $\mu$ and the 
cosmological scale factor,
\begin{equation}
4\pi^2 \ln(a a') i\delta^4(x \!-\! x') + \partial^2 [\tfrac{\ln(\mu^2
\Delta x^2)}{\Delta x^2}] = 4 \pi^2 \ln(\mu^2 a a') i \delta^4(x \!-\! x')
+ \partial^2 [\tfrac{\ln(\Delta x^2)}{\Delta x^2}] \; . \label{connection}
\end{equation}
This suggests that the secular logarithms can be resummed using a variant
of the renormalization group. Because the same two counterterms 
(\ref{DeltaEinstein}) suffice to renormalize any single matter loop 
correction to gravity, a general analysis is possible. In a study of the
massless, minimally coupled scalar loop correction to gravity it was 
shown that a particular combination can be considered as a field strength
renormalization for the graviton \cite{Miao:2024nsz}, 
\begin{equation}
\delta Z = D \Bigl[ 2 (D\!-\!1) c_1 - c_2\Bigr] \kappa^2 H^2 \; . 
\label{deltaZ}
\end{equation}
The same combination works as well for electromagnetism 
\cite{Foraci:2024vng}. Dirac fermions have $c_1 = 0$ and $c_2$ given by 
expression (\ref{c2def}), which implies the gamma function,
\begin{equation}
\gamma \equiv \tfrac{\partial \ln(1 + \delta Z)}{\partial \ln(\mu^2)} =
-\tfrac{\kappa^2 H^2}{160 \pi^2} \; . \label{gamma}
\end{equation}

One can consider both the Weyl tensor (\ref{DiracWeyl}) and the Newtonian 
potential (\ref{DiracNewton}) to be 2-point Green's functions. Hence the 
Callan-Symanzik equation implies,
\begin{equation}
\Bigl[ \tfrac{\partial}{\partial \ln(\mu)} + \beta_G 
\tfrac{\partial}{\partial G} + 2 \gamma \Bigr] G^{(2)} = 0 \; . 
\label{CallanSymanzik}
\end{equation}
The beta function vanishes to the order we are working, and the connection
(\ref{connection}) between $\mu$ and $a$ means that we can replace the
derivative with respect to $\ln(\mu)$ in equation (\ref{CallanSymanzik}) by
a derivative with respect to $\ln(a)$. At this point one sees that the
renormalization group explains the secular factors in expressions 
(\ref{DiracWeyl}) and (\ref{DiracNewton}) and even permits their resummation,
\begin{eqnarray}
C_{0i0j}(x) &\!\!\! \longrightarrow \!\!\!& C_{0i0j}^{\rm tree}(x) \times
[a(t)]^{\frac{\kappa^2 H^2}{80 \pi^2}} \; , \qquad \label{WeylSum} \\ 
\Psi(t,r) &\!\!\! \longrightarrow \!\!\!& \tfrac{G M}{a(t) r} \times 
[a(t) H r]^{\frac{\kappa^2 H^2}{80 \pi^2}} \; . \qquad \label{NewtonSum}
\end{eqnarray}

\section{Conclusions}

We have used dimensional regularization and BPHZ renormalization to 
evaluate the 1-loop contribution of massless Dirac fermions to the graviton 
self-energy (\ref{renormalized}) on an arbitrary cosmological background 
(\ref{background}). Because the massless Dirac Lagrangian (\ref{Dirac}) is 
conformally invariant for any dimension $D$, the primitive contribution 
(\ref{primitive}) is a factor of two times the 2-component result long
ago obtained on flat space background \cite{Capper:1973mv}. The reason we
were able to derive a result for general scale factor $a(t)$ is because 
the scale factor enters only through the counterterms (\ref{DeltaEinstein}). 

We used the Schwinger-Keldysh version of our result (\ref{SKSE}) to 
quantum-correct the linearized Einstein equation (\ref{EinsteinEQN}). That
was solved on de Sitter background to work out 1-loop corrections to the
electric components of the Weyl tensor for plane wave gravitational radiation
(\ref{DiracWeyl}). We also obtained results for the Newtonian potential 
(\ref{DiracNewton}) and the gravitational slip (\ref{DiracSlip}). And we 
were able to resum the results using a variant of the renormalization group
(\ref{WeylSum}-\ref{NewtonSum}).

We could not assume that the primitive contribution from photons would be
the same as in flat space because electromagnetism is only conformally 
invariant for $D=4$ dimensions. This shows up in the photon propagator on
de Sitter background depending on the scale factor and the Hubble constant
for general $D$ \cite{Tsamis:2006gj}. However, the Lagrangian for a 
massless, conformally coupled scalar is conformally invariant for general
$D$,
\begin{equation}
\mathcal{L}_{\rm MCC} = -\tfrac12 \partial_{\mu} \phi \partial_{\nu} \phi
g^{\mu\nu} \sqrt{-g} - \tfrac18 (\tfrac{D-2}{D-1}) \phi^2 R \sqrt{-g} \; .
\label{MMCScalar}
\end{equation}
This means that the primitive contribution to the graviton self-energy 
from a loop of these scalars must agree with the result obtained decades 
previously \cite{Capper:1973bk}. Like our Dirac case, renormalization can
be carried out for an arbitrary scale factor. The Schwinger-Keldysh result
is $\tfrac1{12}$ times that of electromagnetism \cite{Duff:2000mt},
\begin{equation}
[\mbox{}^{\mu\nu} \Sigma_{\rm MCC}^{\rho\sigma}](x;x') = -
\tfrac{\kappa^2}{2^{10} \cdot 3 \cdot 5 \cdot \pi^3} \, 
\mathcal{C}^{\alpha\beta\gamma\delta\mu\nu} \!\times\!
{\mathcal{C}'}_{\alpha\beta\gamma\delta}^{~~~~~\rho\sigma} \Bigl[ 8 \pi 
\ln(a a') \delta^4(x \!-\! x') + f_B(x \!-\! x')\Bigr] . \label{MCCSE}
\end{equation}
In the same way we can solve the linearized Einstein equation 
(\ref{EinsteinEQN}) and resum the results using the renormalization group,
\begin{eqnarray}
C_{0i0j} &\!\!\! = \!\!\!& C_{0i0j}^{\rm tree} \Bigl\{1 \!+\!
\tfrac{\kappa^2 H^2}{480 \pi^2} \, \ln{(a)} + \dots \Bigr\} \longrightarrow
C_{0i0j}^{\rm tree} \times [a(t)]^{\frac{\kappa^2 H^2}{480 \pi^2}} ,
\label{ScalarWeyl} \\
\Psi &\!\!\! = \!\!\!& \tfrac{G M}{a r} \Bigl\{1 \!+\! \tfrac{\kappa^2}{
720 \pi^2 a^2 r^2} \!+\! \tfrac{\kappa^2 H^2}{480 \pi^2} \ln(a H r) \!+\! 
\dots \Bigr\} \longrightarrow \tfrac{G M}{a r} \!\times\! [a(t) H r]^{
\frac{\kappa^2 H^2}{480 \pi^2}} . \qquad \label{ScalarNewton}
\end{eqnarray}

Another interesting spin-off from our work concerns the issue of 
generalizing de Sitter results to a general cosmology. Because the Dirac 
and scalar self-energies (\ref{SKSE}) and (\ref{MCCSE}) are valid for an 
arbitrary scale factor $a(t)$, we can use them to solve the linearized
Einstein equation (\ref{EinsteinEQN}) for a general background. All that is
necessary is to work out the Lichnerowicz operator for a general cosmological 
background and then solve the resulting equation. Even if this cannot be done 
exactly, it can certainly be done numerically.

\vskip .5cm

\centerline{\bf Acknowledgements}

This work was partially supported by NSF grant PHY-2207514 
and by the Institute for Fundamental Theory at the University of Florida.

\end{document}